\documentclass{article}

\usepackage[english]{babel}

\usepackage[letterpaper,top=2cm,bottom=2cm,left=3cm,right=3cm,marginparwidth=1.75cm]{geometry}

\usepackage{ dsfont }
\usepackage{amsmath}
\usepackage{ amssymb }
\usepackage{graphicx}
\usepackage[colorlinks=true, allcolors=blue]{hyperref}
\usepackage{csquotes}
\usepackage{xcolor}

\title{Optimizing Interventions for Agent-Based Infectious Disease Simulations}
\author{Anja Wolpers\textsuperscript{1}, Johannes Ponge\textsuperscript{2}, and Adelinde M. Uhrmacher\textsuperscript{1}\\
[11pt]{
\textsuperscript{1}University of Rostock, Rostock, Germany} \\
\textsuperscript{1}University of M{\"u}nster, M{\"u}nster, Germany\textsuperscript{*} \\
\textsuperscript{*}Also with Stanford University}
\date{}

\begin{document}

\maketitle

\section*{ABSTRACT}
Non-pharmaceutical interventions (NPIs) are commonly used tools for controlling infectious disease transmission when pharmaceutical options are unavailable.
Yet, identifying effective interventions that minimize societal disruption remains challenging.
Agent-based simulation is a popular tool for analyzing the impact of possible interventions in epidemiology.
However, automatically optimizing NPIs using agent-based simulations poses a complex problem because, in agent-based epidemiological models, interventions can target individuals based on multiple attributes, affect hierarchical group structures (e.g., schools, workplaces, and families), and be combined arbitrarily, resulting in a very large or even infinite search space.
We aim to support decision-makers with our Agent-based Infectious Disease Intervention Optimization System (ADIOS) that optimizes NPIs for infectious disease simulations using Grammar-Guided Genetic Programming (GGGP). 
The core of ADIOS is a domain-specific language for expressing NPIs in agent-based simulations that structures the intervention search space through a context-free grammar.
To make optimization more efficient, the search space can be further reduced by defining constraints that prevent the generation of semantically invalid intervention patterns. 
Using this constrained language and an interface that enables coupling with agent-based simulations, ADIOS adopts the GGGP approach for simulation-based optimization. 
Using the German Epidemic Micro-Simulation System (GEMS) as a case study, we demonstrate the potential of our approach to generate optimal interventions for realistic epidemiological models.

\section{Introduction}
\label{sec:intro}
Infectious diseases impose substantial burdens on populations across multiple dimensions.
Beyond the direct health impacts of morbidity and mortality, uncontrolled disease transmission can lead to economic losses through reduced workforce productivity.
When pharmaceutical interventions such as vaccines are unavailable or limited, policymakers must rely on non-pharmaceutical interventions (NPIs), like self-isolation measures and institutional closures, for example, to curb transmission~\cite{geffen_isolation_2020,weigl_household_2021,zhang_evaluating_2022}. 
However, selecting appropriate NPIs presents considerable challenges, as these measures carry their own societal costs. 
Workplace closures disrupt economic activity, school closures compromise educational outcomes for future generations, and prolonged isolation measures adversely affect mental health across populations~\cite{chatterjee_epidemics_2020,chen_metric_2024,giallonardo_impact_2020,jin_economic_2021}.
Consequently, identifying intervention strategies that effectively limit disease spread while minimizing disruption to daily life remains an important challenge for public health decision-making.

Finding balanced NPIs can be assisted by simulating infections and interventions~\cite{matrajt_evaluating_2020,ponge_standardized_2024,prem_effect_2020}.
To this goal, agent-based infectious disease models can simulate complex individual behaviours and social interactions~\cite{lorig_agent-based_2021}.
Thus, agent-based models can replicate regionally heterogeneous demographics and interactions between regions~\cite{ponge_evaluating_2023}.
One example for such an approach is the German Epidemic Micro-Simulation System (GEMS)~\cite{ponge_standardized_2024}, which is an agent-based modeling framework for infectious diseases.
As part of the framework, it is possible to specify NPIs to mitigate the effects of disease outbreaks and evaluate their effectiveness.

Unfortunately, NPIs can be arbitrarily complex, making it difficult to choose which interventions to simulate.
Firstly, interventions can become complex quickly because they often involve combinations of strategies.
For example, during the COVID-19 pandemic, an NPI could entail that a person experiencing symptoms was required to test themselves for antibodies,and depending on the test result, they would be expected to go into self-isolation.
In addition, their school or workplace might be closed for some time to limit the spread of the virus.
So, any combination of viable strategies is part of the search space for the optimal intervention.
Moreover, epidemiological models -- such as GEMS -- are often agent-based for the reasons explained above.
Each agent is typically characterized by several properties and relationships that might constrain or define the applicability and effectiveness of these interventions.
Therefore, an intervention could target only agents whose age attribute is greater than 18, or those who attend school and require them to self-isolate, for example.
Interventions can target individual agents based on the agents' individual properties, as shown above, or the groups to which they belong. 
As such, an intervention could demand that schools be closed whenever one of their pupils becomes infected. 
Thus, these groups reflect different kinds of relationships, such as schools, workplaces, families, or even casual encounters, like meeting at the local grocery store. 
Groups are not exclusive, so each agent may belong to different groups simultaneously, even of the same type, such as two different families. 
Groups can be seen as (temporary) organizational entities with their own identity, properties, and behaviors, which may comprise agents and other groups, thereby enabling a multi-level modeling approach \cite{ponge_targeted_2025,maus_rule-based_2011,Moehring1996}.
Due to this multi-level structure and the possibility of combining interventions arbitrarily, the search space of all possible interventions can become infinitely large, depending on the respective model. 

Usually, people (decision-makers) come up with a potential strategy and modelers use their frameworks to evaluate what the impact of those strategies would be.
However, the results are only as good as the imagination of the decision-makers who devise the strategies. 
Thus, as a first step towards a system that advises decision-makers by automatically generating and evolving complex NPIs, we propose to optimize NPIs for infectious disease simulations using grammar-guided genetic programming.
The Agent-based Infectious Disease Intervention Optimization System (ADIOS) structures the search space of NPIs by defining a language for generating interventions for epidemiological agent-based simulations.
The language is flexible in its core structure, defining the general syntax for intervention composition, while allowing model-specific vocabulary to be added separately.
ADIOS further reduces the search space by excluding semantically invalid intervention patterns by attaching bans to the grammar.
While the grammar ensures syntactic validity, it allows many interventions that are semantically meaningless in the model's context.
For example, repeatedly applying the same filter to a set changes nothing after the first application, and certain combinations of operations may return empty sets given the model's structure (for example, filtering for agents that are both adults and children).
ADIOS therefore provides a ban mechanism that encodes domain knowledge about the model, helping eliminate redundant exploration of functionally identical or pointless interventions and thereby improving the efficiency of the optimization process.

We use the language's grammar and the bans to automatically generate and optimize interventions via grammar-guided genetic programming (GGGP)~\cite{mckay_grammar-based_2010,whigham1995grammatically}.
Genetic programming is a variant of evolutionary algorithms that optimizes a program (often represented by a parse tree with executable tokens as nodes) based on some target function~\cite{koza_genetic_1994}.
In grammar-guided genetic programming, a grammar restricts the search space to a combination of tokens that form syntactically sound (and thus executable) programs.
We leverage this approach to optimize interventions.
In our approach, we represent interventions as parse trees that conform to the grammar's rules and respect the defined bans.
The genetic programming algorithm then evolves a population of interventions over multiple generations, using crossover and mutation operations to explore the search space.
By evaluating interventions through simulation and selecting those that best achieve specified objectives, such as minimizing infections while limiting economic disruption, the algorithm discovers effective intervention strategies that would be difficult to identify manually.
Our approach differs from other grammar-guided genetic programming methods in how we restrict the search space.
While existing approaches learn biases for individual grammar rules~\cite{whigham1995grammatically}, exclude rules based on the optimization task~\cite{hemberg_domain_2019}, or add semantic conditions that abort generation when violated~\cite{de_la_cruz_echeandia_attribute_2005}, we explicitly ban problematic patterns of rule combinations before generation occurs.

We introduce our language for NPIs (including its grammar) in section~\ref{sec:language}.
Section~\ref{sec:additional_restrictions} explains how we reduce the search space of potential interventions by excluding semantically invalid rule expansions using bans.
Then, Section~\ref{sec:geneticProgramming} gives a more detailed overview of how we use the language and the bans in the genetic programming algorithm. 
Finally, we demonstrate our approach in sections~\ref{sec:caseStudy}, and discuss and elaborate on future directions in section~\ref{sec:discussion}.

\section{Components of Non-Pharmaceutical Interventions}\label{sec:NPI}
Non-Pharmaceutical Interventions (NPIs) in agent-based infectious disease simulations are behavioral modifications implemented to reduce disease transmission without the use of vaccines or medications.
In the context of agent-based models, NPIs are represented as programmatic rules that modify agent behaviors, interactions, or movement patterns within the simulation.
In general, we use interventions to define \textit{when}, \textit{which} agents (or groups of agents) are modified \textit{how}.
\textit{Measures} define \textit{how} agents or groups are changed by modifying their property values or relationships, which then may affect their simulated behavior.
For example, a measure could "tell" a group representing a school to shut down by setting its boolean property "is-open" to false.
So, during the simulation, the agents associated with that group would no longer meet and potentially infect each other.

Interventions select \textit{which} agents and groups are targeted by measures using combinations of \textit{filters} and \textit{mappings} applied to the set of all agents or all groups. 
A \textit{filter} applied to a set of agents or groups defines a subset based on the agents' or groups' properties and relationships.
Thus, for example, targeting all agents over the age of 18 or targeting all agents who have a relationship with a group of the type office can be realized by applying filters.  
\textit{Mappings} modify a set of agents or groups by mapping the individual agents or groups to other agents or groups based on their relationships.
This allows for the definition of more complex interventions, such as "All schools where at least one pupil is symptomatic have to close".
This example combines a filter, a mapping, and a measure:
The filter selects all agents with symptoms, the mapping maps from those agents to their school group (if they attend a school), and finally, the measure shuts down these schools.

\begin{figure}[h]
    \centering
    \includegraphics[width=0.6\linewidth]{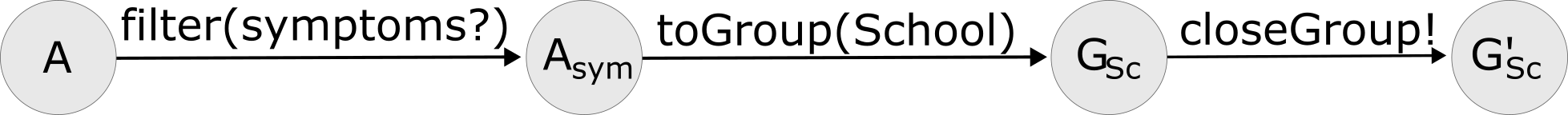}
    \caption{Example intervention tree for the intervention "All schools where at least one pupil is symptomatic have to close".}
    \label{fig:interventionTree-ExampleCloseSchools}
\end{figure}

When dealing with more complex interventions, we represent interventions as trees for clarity.
The tree's nodes represent sets of agents or groups, and a directed edge between two nodes indicates that a filter, mapping, or measure is applied to the starting node, resulting in the goal node.
We call these trees intervention trees.
The intervention tree for the example "All schools where at least one pupil is infected have to close" is visualized in Figure~\ref{fig:interventionTree-ExampleCloseSchools}.
The first node, $A$, is the set of all agents.
The outgoing edge from this node is the filter $filter(symptoms?)$.
It filters all agents for those who have symptoms resulting in node $A_{sym}$.
The mapping operator $toGroup$ maps from a set of agents to a set of groups. 
So, with the argument $school$, it checks each agent in the group of infected agents for relationships with groups of the type "school". 
Only these groups are then elements of the resulting set $G_{Sc}$ that represents the schools where at least one pupil is infected. 
The set $G_{Sc}$ is the target for the measure $closeGroup!$ that sets all the groups' is-open properties to false.
This creates the set $G'_{Sc}$ that contains the same but modified groups as $G_{Sc}$.
Note that this combination and sequence of operations is not the only way this intervention can be realized. 
For example, one could filter for children before filtering for symptomatic agents, and still achieve the same effect (but possibly the runtime might differ). 

\begin{figure}[h]
    \centering
    \includegraphics[width=0.6\linewidth]{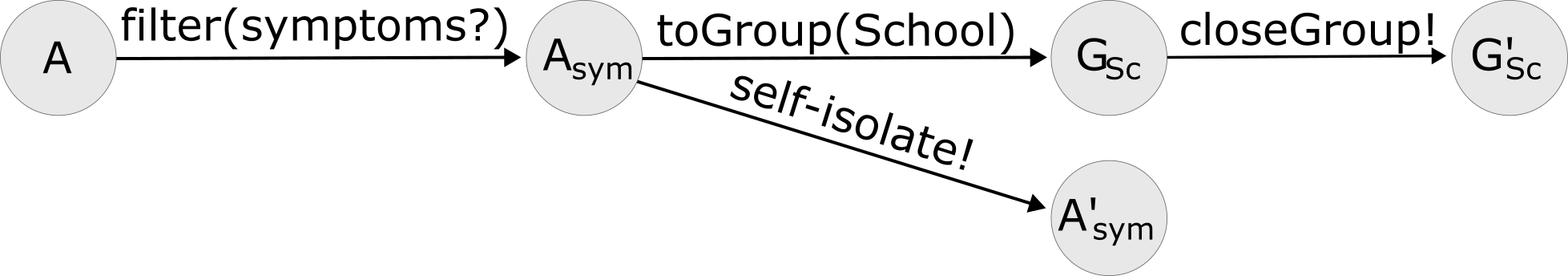}
    \caption{Example intervention tree for the intervention "Symptomatic students have to self-isolate, and their school has to close".}
    \label{fig:interventionTree-ExampleCloseSchoolsAndSelfIsolate}
\end{figure}

As mentioned in the introduction, interventions can be combined. 
Our intervention trees represent this by allowing multiple outgoing edges per node. 
Consequently, multiple measures can be applied to the same set of agents or groups and intermediate sets.
Moreover, intermediate sets can be transformed by multiple filters and mappings simultaneously. 
This allows for interventions like "Symptomatic students have to self-isolate, and their school has to close" (see Figure~\ref{fig:interventionTree-ExampleCloseSchoolsAndSelfIsolate}).
Finally, triggers define \textit{when} interventions are executed. 
Interventions can be executed at each tick during the simulation, or an intervention might be triggered by a condition becoming true, e.g., more than 100 agents being infected.
Based on this structure of interventions, we will introduce a language for interventions for agent-based infectious disease simulations, including (context-free) grammar in EBNF.

\section{Our Domain-Specific Language for Non-Pharmaceutical Interventions}\label{sec:language}
The domain-specific language (DSL) at the heart of ADIOS formalizes and generalizes the previously described structure of interventions, which consists of a trigger, filters, mappings, and measures applied to sets of agents or groups.
It can be adapted to specific models flexibly and only defines how triggers, filters, mappings, and measures for agents and groups can be combined.
Thus, model-specific triggers, measures, mappings, and filters must be set separately.
It assumes that a set of agents (A) and some higher-order grouping (G) exist within the agent-based model.
The agents and groups have relationships with each other and properties whose values can be requested and assigned.

\textit{Filters} operating on a set of agents are represented by the token $a>a(p_a?)$.
The $p_a?$ part of this token is a placeholder for an expression that the set of agents is filtered for, and depends on the specific model. 
So, using the agents' property values, one can formulate model-specific expressions $p_a? \in P_a?$ that can be evaluated to be either true or false, e.g., whether an individual has symptoms.
\textit{Filters} for sets of groups work just like filters for sets of agents. 
They are represented by the token $g>g(p_g?)$ where $p_g?$ must be substituted by a boolean expression $p_s? \in P_s?$ about a group's properties, like whether a group has more than 30 employees.
\textit{Measures} for agents ($a!(p_a!)$) and groups ($g!(p_g!)$) require further definitions of $p_a! \in P_a!$ and $p_s! \in P_s!$ denoting how property values of an agent or of a group are changed.
These changes might be, for example, an individual moving into self-isolation or a school being closed, respectively. 
\textit{Mappings} use relations within the set of agents and groups, as well as between agents and groups. 
Thus, our language differentiates between different kinds of mappings.
For instance, mapping from a set of agents to another set of agents has the token $a2a(r_{a,a})$ where $r_{a,a} \in R_{a,a}$ specifies how agents from the first set of agents are related to agents in the second one. 
For example, $r_{a,a}$ could specify a mapping representing tracing the infectious contacts of infected agents.
So, for each infected agent in the set that the mapping is applied to, the agents that they infected are added to the goal set, resulting in one set of agents containing all agents that have been infected by any agent from the first set.
All other mappings work similarly: the relations within the model ($r_{a,g} \in R_{a,g}$, and $r_{g,a} \in R_{g,a}$) form the basis for mapping a set of agents to a set of groups ($a2g(r_{a,g})$), and vice versa ($g2a(r_{g,a})$), and relations within the set of groups ($r_{g,g}\in R_{g,g}$) allow mapping from a set of groups to another set of groups ($g2g(r_{g,g})$).
All these operators can be combined based on whether they are operating on sets of agents and groups. 
So, operators' outputs must match the type of set that the following operators are operating on, as, for example, it makes no sense to close an individual, but one can close a school.

We formalize our intervention language's syntax using an extended context-free grammar.
Regular context-free grammars are a four-tuple $<N,\Sigma, P, S>$ consisting of a non-terminal alphabet N, a terminal alphabet $\Sigma$, a set of production rules, and a start symbol S.
For context-free grammar, we split the non-terminal alphabet, the terminal alphabet, and the production rules into two non-terminal alphabets (N and H), two terminal alphabets ($\Sigma$ and $\tau$), and two sets of production rules (P and Q).
N, $\Sigma$, and P are the non-terminal alphabet, the terminal alphabet, and production rules for our language. 
$\tau$ and Q are the terminal alphabet and production rules that have to be added by the user to make the language work for their model-specific application.
The hybrid alphabet H functions as an alphabet of terminals for the general part of our language, but its symbols are non-terminals for the model-specific language, i.e., the symbols in H are the left-sides of the rules in Q.
Thus, the extended context-free grammar for our language is the seven-tuple $<N,H,\Sigma,\tau, P, Q, S>$.
The start symbol S is \textit{intervention}, and the non-terminal alphabet N contains the start symbol and the symbols starta and startg that define whether the next operation is working on the set of agents (starta) or the set of groups (startg) that are part of the current state of the model. 
\def\t#1{\textcolor{gray}{"#1"}}
\def\h#1{\textcolor{teal}{#1}}
For brevity, the symbols of the terminal alphabet $\Sigma$ and the hybrid alphabet H are only highlighted in the first set of production rules P, below. 
The symbols of the terminal alphabet $\Sigma$ are \textcolor{gray}{gray} in the symbols of the hybrid alphabet H are \h{green}.

\begin{equation}
    intervention = \t{[}\h{trigger}\t{]}\ (starta)\ |\ (startg)
    \label{eq:intervention}
\end{equation}
\begin{equation}
    \begin{split}
        starta =\ 
            &(\t{(}starta\ \t{\&}\ starta\t{)})\ | \\
            &(\t{a>a(} \h{p_{a}?}\t{).}starta)\ | \\
            &(\t{a2a(}\h{r_{a,a}}\t{).}starta)\ | \\
            &(\t{a2g(}\h{r_{a,g}}\t{).}startg)\ | \\
            &(\t{a!(}\h{p_a!}\t{)})
    \end{split}
    \label{eq:starta}
\end{equation}

\begin{equation}
    \begin{split}
        startg =\ 
            &(\t{(}startg\ \t{\&}\ startg\t{)})\ | \\
            &(\t{g>g(} \h{p_{g}?}\t{).}startg)\ | \\
            &(\t{g2a(}\h{r_{g,a}}\t{).}starta)\ | \\
            &(\t{g2g(}\h{r_{g,g}}\t{).}startg)\ | \\
            &(\t{g!(}\h{p_g!}\t{)})     
    \end{split}
    \label{eq:starts}
\end{equation}

The second set of terminals $\tau$ and the set of production rules Q are specific to the model to which our method is applied. 
For example, below is a subset of the production rules Q (with terminals $\in \tau$ highlighted in \textcolor{gray}{gray}) for a GEMS model (that will also be used in the case study in Section~\ref{sec:caseStudy}).
The following rules are needed to express the example from Figure~\ref{fig:interventionTree-ExampleCloseSchoolsAndSelfIsolate}. 
\begin{equation}
    \begin{split}
        \h{trigger} = \t{true}
    \end{split}
    \label{eq:trigger}
\end{equation}
\begin{equation}
    \begin{split}
        \h{p_{a}?} =\ 
        & \t{symptoms?}\ |\ \t{child?}\ |\ \t{adult?}\ |\ \t{positive\_test?}
    \end{split}
    \label{eq:pa}
\end{equation}
\begin{equation}
    \begin{split}
        \h{r_{a,g}} =\ 
        &\t{School}\ |\ \t{Office}\ |\ \t{Household}
    \end{split}
    \label{eq:rag}
\end{equation}
\begin{equation}
    \begin{split}
        \h{p_{a}!} =\ &\t{self-isolate!}
    \end{split}
    \label{eq:pa!}
\end{equation}
\begin{equation}
    \begin{split}
        \h{p_{g}!} =\ &\t{close!}
    \end{split}
    \label{eq:pg!}
\end{equation}
 
\begin{figure}[h]
    \centering
    \includegraphics[width=0.6\linewidth]{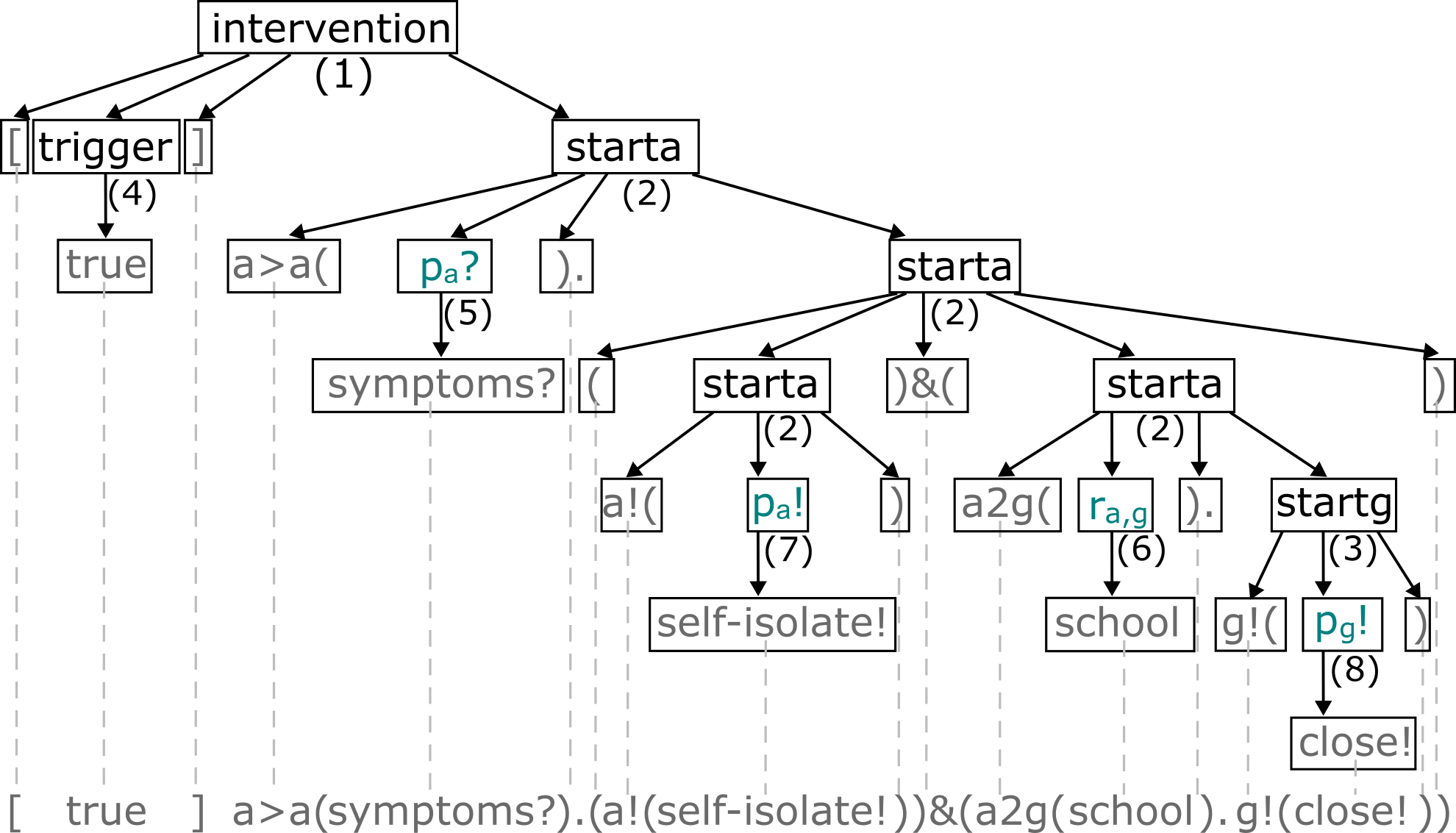}
    \caption{The parse tree for the example intervention tree in Figure~\ref{fig:interventionTree-ExampleCloseSchoolsAndSelfIsolate}. In the brackets are the rules that are applied to the token above to reach the token below.}
    \label{fig:parse-tree}
\end{figure}

Note that the terminals $\in \tau$ are only tokens made up for this language and do not directly interact with the model. 
To generate executable experiments that implement an NPI defined in this language, we require a parser that translates expressions from our NPI language into code that interacts with the model (see Section~\ref{sec:geneticProgramming}). 
Using the (extended) context-free grammar, ADIOS generates interventions by repeatedly applying rules from the sets P and Q to all non-terminal and hybrid symbols until only terminal symbols $\in \Sigma \cup \tau$ are part of the intervention string~\cite{whigham1995grammatically}.
The example intervention from Figure~\ref{fig:interventionTree-ExampleCloseSchoolsAndSelfIsolate} is constructed starting with the start symbol $intervention$, substituting it with $\t{[}\h{trigger}\t{]}\ (starta)$ using rule~\ref{eq:intervention}. 
Then, $trigger$ is substituted by applying rule~\ref{eq:trigger}, and $starta$ is extended by applying rule~\ref{eq:starta}.
Finally, after recursively applying production rules until all symbols are terminals, the intervention string is \textcolor{gray}{"$[true] a>a(symptoms?).(a!(self-isolate!)\&(a2g(School).$ $g!(close!))$"} which only consists of terminal symbols $\in \Sigma \cup \tau$.
All rule applications that are required to reach this intervention string are visualized as a parse tree in Figure~\ref{fig:parse-tree}. 

\section{Search Space Restriction Using Bans} \label{sec:additional_restrictions}
\begin{figure} [h]
    \centering
    \includegraphics[width=0.6\linewidth]{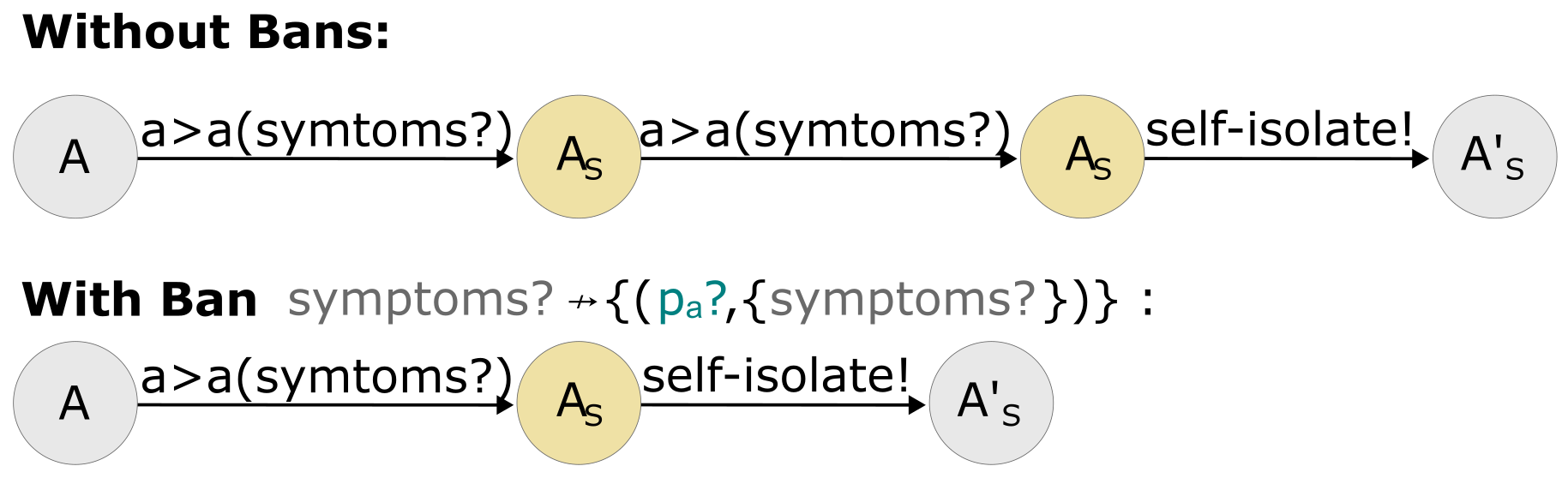}
    \caption{Filtering a set repeatedly with the same filter will repeatedly result in the same set (here, $A_S$).
    Banning repeating identical filters prevents these semantically pointless interventions.}
    \label{fig:interventionTreerepeatedFilters}
\end{figure}

The language from Section~\ref{sec:language} defines the search space of syntactically valid interventions.
However, based on its grammar, the search space includes many interventions with redundant operators. 
For example, it is syntactically correct to apply the same filter multiple times in a row (see Figure~\ref{fig:interventionTreerepeatedFilters}). 
But, semantically, all these filters but the first one do not affect the intervention's effect.
This introduces a bias whereby the presence of multiple semantically equivalent interventions in the search space increases the likelihood that the generator will sample from this overrepresented region, thereby reducing exploratory diversity.
So, ideally, the search space would only consist of semantically unique interventions.
Therefore, we extend our language with constraints that ban patterns of operations leading to semantically equivalent NPIs.

For example, a ban preventing the first intervention from Figure~\ref{fig:interventionTreerepeatedFilters} could be to ban a\textgreater a(symptoms?) being followed by another a\textgreater a(symptoms?).
Thus, when selecting the next operation following the first a\textgreater a(symptoms?), the ban would make ADIOS choose a different operation than a\textgreater a(symptoms?).

We define a ban to be $$l\nrightarrow\{(h,\{t' | t' \in \tau \cup\{previous, self\}\})|h\in H\}, l\in \tau \cup H,$$ where previous and self are optional variables that allow defining abstract bans.
When a ban is applied to an edge in an intervention tree, the variable previous is resolved to a set of terminals $\in \tau$ that were already banned from the bans attached to the previous edge, and the variable self is resolved to the terminal $\in \tau$ that is associated to the current intervention tree edge.
We will demonstrate this in examples below.

A simple ban that prevents repeatedly filtering for symptomatic agents in a row is $$\t{symptoms?} \nrightarrow \{(\h{p_a?},\{\t{symptoms?}\})\}.$$
However, any kind of (deterministic) filter should not be repeated on the same set of agents.
Thus, an abstract ban that prevents repeatedly filtering agents for any attribute in a row is $$\h{p_a?} \nrightarrow \{(\h{p_a?},\{self, previous\})\}.$$

\begin{figure}
    \centering
    \includegraphics[width=0.4\linewidth]{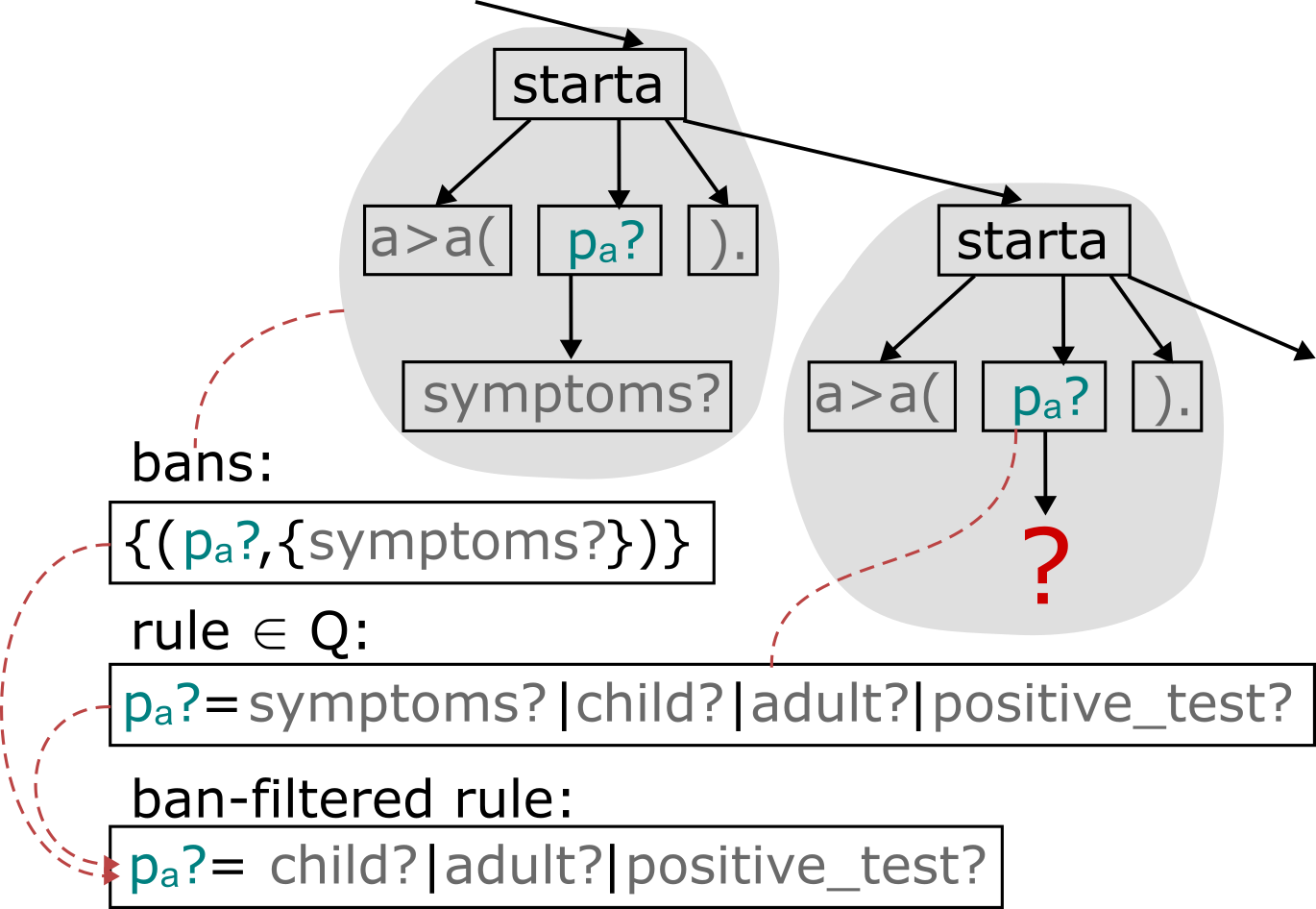}
    \caption{Example of how the bans impact the expansion of symbols: Before the rule~\ref{eq:pa} $\in Q$ for $\h{p_a?}$ is applied, it is filtered by the bans attached to the previous intervention tree edge. An intervention tree edge corresponds to a parse tree's sub-tree starting with a node labeled with starta and ending before the next node labeled starta. Here, the nodes corresponding to the two intervention tree edges are grouped in the gray bubbles.}
    \label{fig:apply-bans-example}
\end{figure}

Using this example ban, we will now go through how the combination of bans and the grammatical rules works when generating interventions with ADIOS. 
Let us assume that the generation of an intervention is at the point visualized by the parse tree in Figure~\ref{fig:apply-bans-example}.
Previously, the node starta has been expanded to the nodes $\t{a>a(}$, $\h{p_a?}$, \t{).}, and starta (using rule~\ref{eq:starta}), and $\h{p_a?}$ has been replaced by \t{symptoms?} following rule~\ref{eq:pa}.
The application of this last rule completes the sub-tree that is represented as the first edge labeled "a\textgreater a(symptoms?)" in the NPI's intervention tree in Figure~\ref{fig:interventionTreerepeatedFilters}.  
At this point, a set of bans is created and attached to this set of nodes representing the intervention tree edge.
The contents of this set of bans are determined by the hybrid token $h\in H$ that is a label in this set of nodes.
In this example, the token $h \in H$ in the set $\{starta, \t{a>a(}, \h{p_a?}, \t{).}, \t{symptoms?}\}$ is $\h{p_a?}$.
So, the bans for $\h{p_a?}$, which are defined above, are resolved by substituting the variable self with the terminal $\t{symptoms} \in \tau$ that is part of the intervention tree edge's node set.
For this example, we will ignore the variable previous and explain its effects below.
Consequently, the set of bans attached to the intervention tree edge is $\t{symptoms?} \nrightarrow\{(\h{p_a?},\{\t{symptoms}\})\}$ (or in short: $\{(\h{p_a?},\{\t{symptoms}\})\}$).

Also, the starta node from the previous expansion has, in turn, been expanded into \t{a\textgreater a(}, $\h{p_a?}$, \t{).}, and starta by applying rule~\ref{eq:starta} again.
Now, the next step is to expand the second $\h{p_a?}$.
However, instead of directly applying rule~\ref{eq:pa}, we filter the rule according to the bans attached to the previous intervention tree edge, which indeed contain bans for expanding the symbol $\h{p_a?}$.
Therefore, the terminal symbol from the ban, \t{symptoms?}, is filtered from rule~\ref{eq:pa}.
The now ban-filtered rule $\h{p_{a}?} =\ \t{child?}\ |\ \t{adult?}\ |\ \t{positive\_test?}$ can then be applied to the hybrid symbol $\h{p_a?}$ without running the risk of repeating the same filter in a row.

In addition to banning specific terminals, the variable previous allows carrying over bans from the previous edge. 
Because, for example, if there was another filter after a first filter filtering for children (a\textgreater a(child?)) that filtered for agents with symptoms (a\textgreater a(symptoms?)), then applying a second filter a\textgreater a(child?) would still result in the same set of symptomatic agents who are children that had already been the result of the a\textgreater a(symptoms?) filter.
The abstract ban for filters defined above prevents this unnecessary repetition of filters. 
So, when generating the intervention edge for a\textgreater a(symptoms?) is finalized, the variable previous makes ADIOS copy the bans (for \h{$p_a?$}) from the previous edge (a\textgreater a(child?)). 
Thus, the bans for the edge following a\textgreater a(symptoms?) are effectively $$\{(\h{p_a?},\{\t{symptoms?},\t{child?}\})\}$$ which forbids choosing a\textgreater a(child?).
This example shows that whenever our approach has added a terminal $t \in \tau$ to an intervention using a rule $q\in Q$, it updates the set of legal picks for follow-up terminals, overwriting the grammar. 

Moreover, depending on the model, there might be syntactically sound interventions that make no sense in the context of the model and can be excluded from the start to save resources.
Therefore, we allow users to formulate bans that exclude pointless interventions from the search space.
For example, it would be a syntactically valid intervention for our GEMS language to close households (instead of closing schools or offices).
However, closing a group makes agents stay at home, which they cannot do if their household is closed, making the intervention pointless.
Model-specific bans like $$\t{Household}\nrightarrow\{(\h{p_a!},\{\t{close-group!}\})\}$$ can prevent such interventions.
Note that such bans are only feasible for models with only a few types of groups (such as household, school, and office).
Banning all operations that are not defined for each group type for a more complex model can become exponentially cumbersome and would, thus, require a group-type-aware version of our language.

\section{Grammar-Guided Genetic Programming to Search for Optimal Interventions}\label{sec:geneticProgramming}
Genetic programming (GP) generates and optimizes a program inspired by natural selection in nature~\cite{koza_programming_1993,koza_genetic_1994,sobania_comprehensive_2023}. 
Initially, a population of programs (also called individuals) is generated and evaluated. 
Then, new generations are created based on the individuals of the previous generation.
The programs in the previous generation are selected for cloning, mutation, or crossover based on their fitness, and thus their impact on the new generation depends on the evaluation. 
Thus, the overall fitness of the population and of the individual programs increases over time. 
Common hyperparameters of genetic programming algorithms are the size of the population, the number of generations, and the percentage of crossovers and mutations. 

Since its inception, (context-free) grammars have been used to restrict GP's search space by syntactically formalizing the structure of viable solutions~\cite{koza_programming_1993, mckay_grammar-based_2010}. 
This field of GP has since been termed grammar-guided genetic programming (GGGP) and is a widely used variant of GP.
GGGP has been applied to a variety of domains, including ecology~\cite{mckay_variants_2001}, image segmentation~\cite{herrera-sanchez_multiple-feature_2025}, traffic planning~\cite{keblawi_grammatical_2025}, engineering~\cite{danandeh_mehr_genetic_2018}, and robotics~\cite{tanev_genetic_2007}.
Some approaches restrict their search spaces further by learning biases for their grammar's rules~\cite{whigham1995grammatically}, by removing grammar elements based on the optimization task~\cite{hemberg_domain_2019}, or by adding semantic information to the grammar and aborting the generation of semantically wrong programs~\cite{de_la_cruz_echeandia_attribute_2005}.

Genetic programs generally have two representations, called genotype and phenotype~\cite{mckay_grammar-based_2010}.
The genetic algorithm operates on the genotype, which includes generating, mutating, and crossing programs. 
In GGGP, the individual programs' genotypes are commonly parse trees~\cite{whigham1995grammatically} or linear strings~\cite{oneil_grammatical_2003}.
The phenotype is the representation that is executed for evaluation, which is an expression tree for most GGGP approaches.

\begin{figure}
    \centering
    \includegraphics[width=0.5\linewidth]{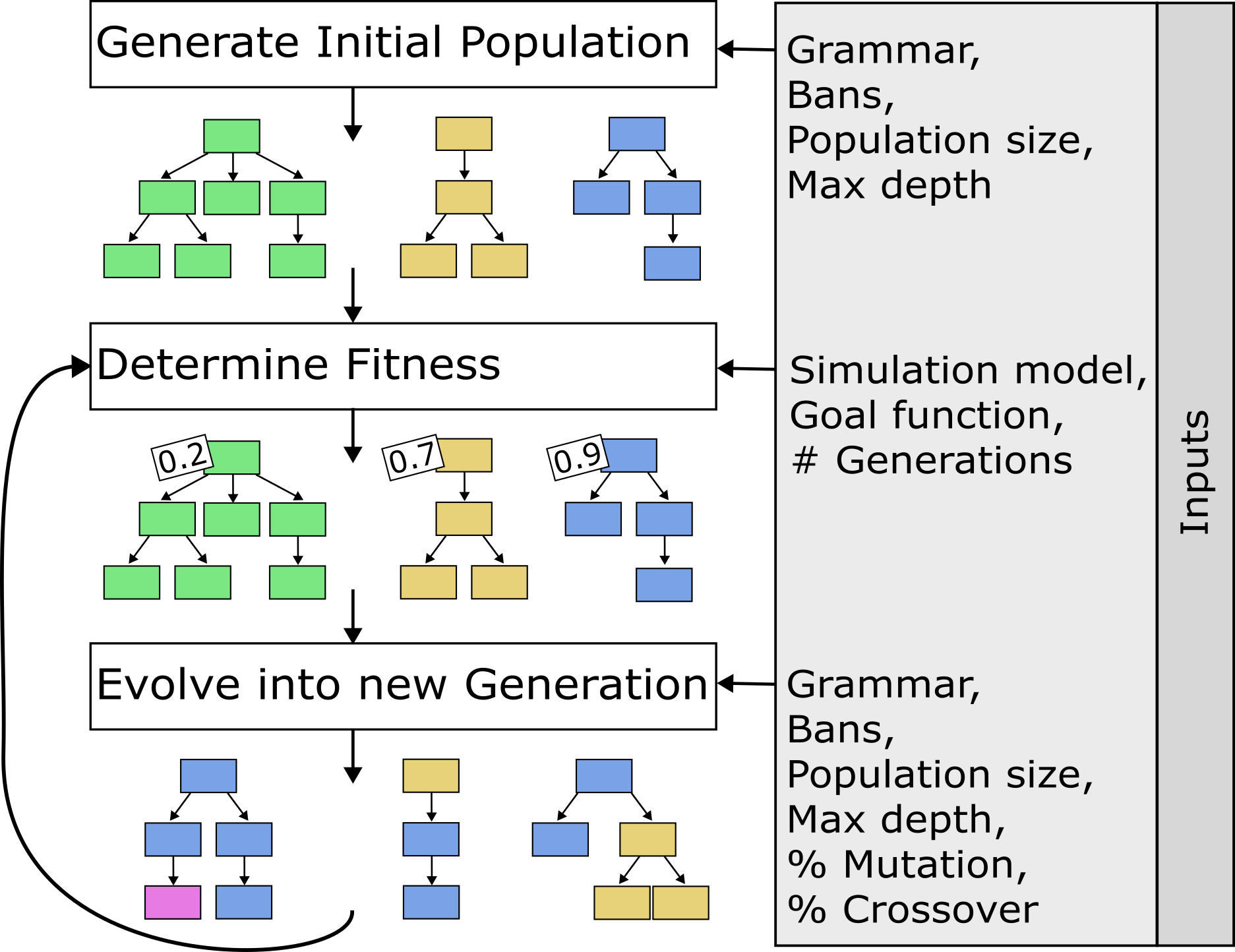}
    \caption{The steps in our GGGP algorithm. 
    First, a new generation of NPIs is generated based on our language's grammar. 
    Then, the NPIs are evaluated in the context of a given goal function. 
    Next, based on the NPIs' fitness values, they are evolved into a new generation.
    The green intervention has a low fitness and is therefore not part of the new generation. 
    The blue one has the best fitness and is cloned into the next generation, then crossed with the yellow one. 
    In addition, one intervention in the new generation has been mutated, which is visualized by the pink node that has not yet been part of the population.}
    \label{fig:geneticAlgorithm}
\end{figure}

We use GGGP to generate and optimize non-pharmaceutical interventions for infectious disease simulations in the language described in section~\ref{sec:language}. 
In our approach, the genotypes are parse trees in our language (see the example in Figure~\ref{fig:parse-tree}) that span the search space.
In addition to our context-free grammar, we limit the search space by banning patterns of intervention tree edges using the domain knowledge of the users who know their model (see Section~\ref{sec:additional_restrictions}). 
We evaluate an NPI's fitness by executing an experiment specification (in a programming language) that is generated from the NPI string (in our NPI language) consisting of the NPI parse tree's leaves.
Thus, the phenotype is the experiment generated from the NPI. 
In the following, we will go through how our approach uses the GGGP algorithm in more detail, following the three steps illustrated in Figure~\ref{fig:geneticAlgorithm}.

\paragraph{Step 1: Generating the initial population.}
First, our approach randomly generates an initial NPI population of a given size. 
The interventions' genotypes are generated in the form of parse trees by repeatedly applying the grammar's rules while also adhering to the respective bans as shown in the example in Section~\ref{sec:additional_restrictions}. 
The resulting parse trees' depths are limited by a max-depth parameter, thus avoiding infinitely large interventions. 
These interventions' phenotypes are generated by merging the parse trees' leaves from left to right, creating string expressions in our language (see Section~\ref{sec:language} for an example).

\paragraph{Step 2: Determine fitness of candidates.}
The NPIs in the population are then evaluated by running simulations with the model and computing their fitness according to a given goal function.
In general GGGP, the generated genotypes and phenotypes are expression trees of complete programs that are directly executed to evaluate their fitness. 
In contrast, our approach generates NPIs in our domain-specific language that are not executable on their own.
Therefore, an extra step is required to map the generated genotype (the NPI parse-tree) to an executable phenotype in the form of an experiment specification in a programming language that interacts with the simulation model during runtime, enforcing the NPI.
This experiment specification must implement the generated NPI, execute the simulation model, and observe the model parameters needed to evaluate the goal function.
Running the generated experiments and evaluating the goal function for each run returns the fitness score of the respective NPI.
The fitness score reflects how well the intervention contributes to maximizing the optimization goal and, thus, influences its impact on the generation of the following generations.

If the target number of generations as part of the algorithm's hyperparameters has been reached at this point, the most fit NPI is returned as the solution of the optimization heuristic.

\paragraph{Step 3: Evolve into a new Generation.}
Evolving the previous generation of NPIs into a new one closely follows the evolution of programs in GGGP~\cite{mckay_grammar-based_2010,whigham1995grammatically}.
Based on the individuals' fitness scores, the new generation is generated by cloning, mutating, and crossing individuals from the previous generation.
The main difference to other GGGP methods is that our versions of mutation and crossover have to adhere to the bans in addition to the grammar. 

The most fit individual is cloned and added to the new generation without changes. 
All other NPIs in the new generation are generated by selecting the most fit two individuals from a random subset of individuals from the previous generation (i.e., tournament selection~\cite{talbi_metaheuristics_2009}).
Depending on the GGGP's hyperparameters controlling the percentage of cross-overs and mutations, the selected interventions undergo cross-over and/or mutation before they are added to the new generation.

\begin{figure}
    \centering
    \includegraphics[width=0.7\linewidth]{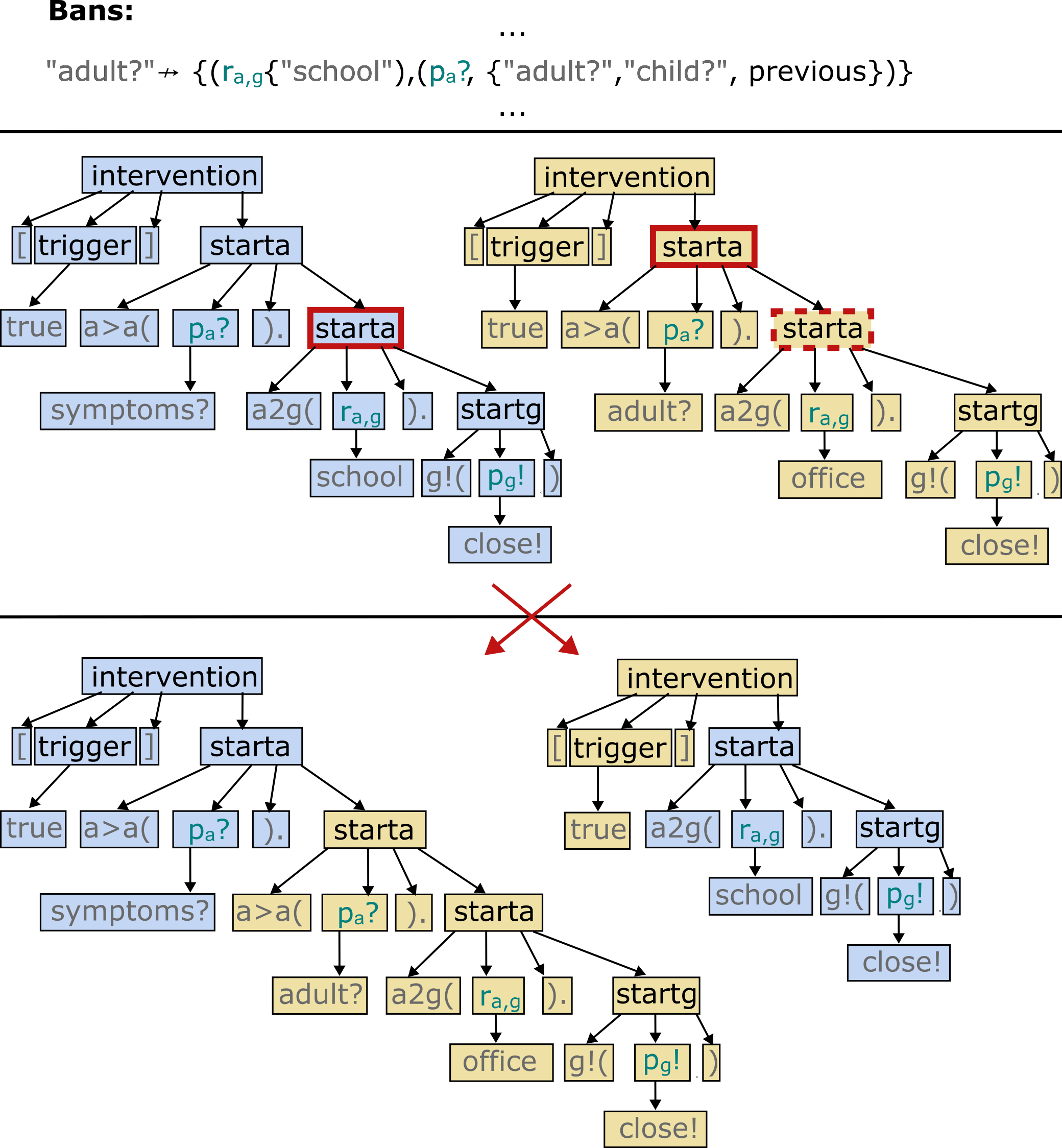}
    \caption{Example for a crossover between two intervention parse trees.
    With the given ban, the highlighted starta node in the left tree and the starta node in the right tree with the dashed outline are not a viable option for crossover.
    However, the combination of the left starta node and the other highlighted right starta node with the continuous outline is a legal crossover pair.}
    \label{fig:crossover}
\end{figure}

If two individuals are selected for crossover, then elements from two parent programs are combined to create offspring.
Like in crossover for regular GGGP, crossover points with two non-terminal nodes with the same labels are selected from both parse trees~\cite{mckay_grammar-based_2010}.
Our approach traverses potential crossover points and checks whether they are also compatible regarding bans.
Then, one of the filtered crossover points is chosen at random, and the respective nodes, including their sub-trees, are swapped between the two parse trees, thereby creating two new and different parse trees that become part of the next generation. 
Figure~\ref{fig:crossover} shows an example of a crossover between two interventions. 
When filtering all potential crossover pairs, our algorithm goes through all pairs of nodes with the same label (each node in the pair stems from one of the two parse trees).
For this example, we only look at two potential crossover points consisting of the highlighted nodes labeled starta: the only highlighted node from the left tree and either the node with the dashed or continuous outline in the right tree.
The node with the dashed outline is not a valid crossover partner for the node from the left tree because the ban for this example forbids that a "adult?" filter is followed by a mapping to "school" groups, which would be the case after this crossover. 
In contrast, crossing the two interventions at both nodes with continuous outlines does not violate any bans.

During mutation, a sub-tree is newly generated starting from a randomly selected non-terminal node in the individual's parse tree.
Thus, a node in the intervention's parse tree is selected, and all its child nodes are deleted.
Starting from this node, the grammar's rules are applied while satisfying the bans until the max-depth is reached or all expressions at the tree's leaves are terminal, as during the generation of the initial population in step 1.
Mutation ensures that later generations still explore the search space, trying out expressions that have not yet been part of the population.
 
\paragraph{Repeat steps 2 and 3.}
Newly generated populations of NPIs are again evaluated as described in step two. 
Since the new generation mainly contains interventions generated from the best performing interventions from the last generation, the population's overall fitness should improve for each generation.
After a previously defined number of iterations, no new generation is generated, and the best performing intervention is returned.

\section{Case Study: GEMS Interventions}\label{sec:caseStudy}
We demonstrate our approach in practice by applying it to a simple GEMS simulation model of its default pathogen that is loosely based on COVID-19 (using GEMS version v.0.7.0).
Our chosen model simulates infections using GEMS' population model of the Saarland, the smallest German area state with about a million inhabitants.
The individuals are associated with groups (called "settings" in GEMS) that simulate households, schools, workplaces, and municipalities according to realistic demographic structures~\cite{gesylandeu}.
Note that these experiments are of a purely hypothetical nature and designed to test ADIOS. 
All results are thus only to be interpreted in the context of testing ADIOS and cannot be generalized.

\subsection{ADIOS Setup for GEMS}
As model-specific terminal tokens for our language ($\tau$), we use the mappings and measures predefined in the GEMS framework. 
The predefined mappings in GEMS allow mapping from a set of groups to the set of agents in any of the groups (\t{members}), from a set of agents to a set of groups of a given type that any of the agents is associated with (\t{office}, \t{school}, \t{household}), and from a set of agents to the set of agents that have been infected by an agent of the previous set (\t{trace\_infectious\_contacts}). 
GEMS includes three predefined measures: it is possible to close and reopen a group, and to send agents into self-isolation for a given number of days, starting on the day after the intervention is issued. 
Based on GEMS' measures, we defined the measures \t{self-isolate-14}, \t{self-isolate-7}, and \t{close-group-7}.
The measures \t{self-isolate-14} and \t{self-isolate-7} encode that the GEMS measure self-isolation is executed with the parameter for the duration set to 14 or 7, respectively.
Similarly, the measure \t{close-group-7} combines the GEMS measures to close and open groups so that a group is closed for 7 days starting from the day the intervention is issued.
GEMS includes only one predefined filter, \t{positive\_test?}, which filters agents with positive antigen test results, but more filters can be defined based on agent and group parameters. 
Therefore, we defined the following filters for agents: \t{symptoms?}, \t{child?}, \t{adult?}, and \t{older\_than\_67?}, and \t{more\_than\_20\_members?} and \t{at\_most\_20\_members?} as filters for groups.
For the moment, the only terminal token for a trigger in our $\tau$ is \t{true}.
That is, because the predefined triggers in GEMS are indirectly already part of the interventions.
Most prominently, GEMS defines a symptom trigger that issues an intervention to all agents with symptoms. 
Our intervention language for GEMS can model such interventions without this trigger by using the \t{symptoms?} filter.
In addition, GEMS contains triggers (one for agents and one for groups) that start an intervention each tick (= day) in a given tick interval.
We use these triggers by default when no trigger is given (i.e., the trigger is the terminal \t{true}). 

Using the GEMS Saarland model and our intervention language with the model-specific extensions explained above, we search for optimal NPIs that minimize the simulated disease's impact on the economy, education, and public health.
Thus, our goal functions are based on the number of lost work days for the average working person ($lost\_work\_days$), the number of lost school days for the average student ($lost\_school\_days$), and the average number of days that an individual is infected ($sick\_days$).
With this setup, we generated and optimized NPIs for four scenarios with differing goal functions.

The first three scenarios each aim to minimize one of the above parameters in isolation, allowing us to examine if the resulting NPIs can plausibly minimize the respective parameter and thus provide validation for our approach, since there are no known optimal NPIs that we can use for validation.
The fourth scenario optimizes for a combined goal function that balances all three parameters by minimizing their weighted sum: $lost\_work\_days + lost\_school\_days + 2 * sick\_days$.
We weighted the sick\_days by a factor of 2 to reflect that the primary goal of our policy is to limit infections.

For each scenario, we used a population of ten parse trees with a maximum depth of five that were evolved over four generations. 
During evolution, individuals from the previous generation are selected using a tournament selection of size 3, undergo crossover with probability 0.7, and are mutated with probability 0.5. 
Each intervention was evaluated by running the GEMS simulation over a simulated period of one year.

\subsection{Results}
\begin{figure}
    \centering
    \includegraphics[width=0.5\linewidth,trim={0cm 1.5cm 1cm 2cm}]{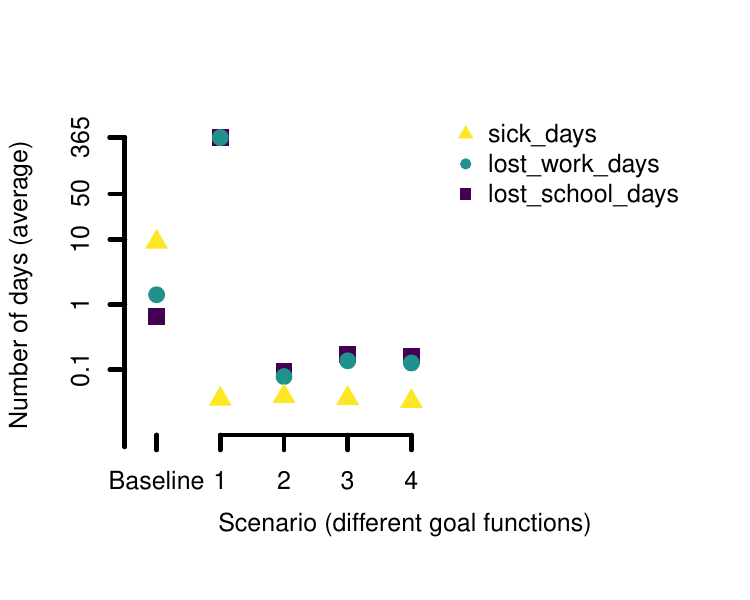}
    \caption{Observed Parameter values for the fittest NPI per scenario and for a baseline simulation without any NPIs.
    Scenario 1: Minimize sick days; Scenario 2: Minimize lost school days per student; Scenario 3: Minimize lost work days per worker; Scenario 4: combined goal function (lost\_work\_days + lost\_school\_days + 2 * sick\_days).}
    \label{fig:ParameterValuesPerScenario}
\end{figure}

Scenario one aims to minimize sick\_days.
The best performing solution generated by ADIOS was the NPI [true] a!(self-isolate-14).
This intervention achieved approximately 0.04 sick days per person in the simulated year, representing a reduction of 9.11 days compared to the baseline scenario without interventions, with approx 9.15 sick days per person (see Figure~\ref{fig:ParameterValuesPerScenario}).
As this NPI contains no filter selecting agents based on their attributes, all agents are sent into self-isolation (for 14 days) at each tick in the simulation.
Even though it may not be intuitive that the agents stay in isolation for longer than 14 days, this intervention has the effect that all agents spend the whole simulation in perpetual self-isolation.

The second scenario aims to minimize \textit{lost\_school\_days} (the days that students cannot attend school).
The fittest NPI for this scenario was [true] a\textgreater a(positive\_test?) .a!(self-isolate-14) with 0.095 missed school days per student on average over the simulated year, improving the average 0.67 missed school days in the baseline.
This NPI does not require all agents to go into self-isolation; only those with a positive antibody test (which is 100\% accurate in our implementation). 
This also has a positive effect on the average number of missed work days (0.088) and sick days (0.04), even though these parameters were not part of the goal function.
Thus, this NPI slightly outperforms all NPIs found for the following scenarios.

In the third scenario, we sought an NPI to minimize the number of days workers are unable to work ($lost\_work\_days$), which resulted in the rather long NPI [true] (a\textgreater a(positive\_test?). a2g (Household). g\textgreater g(at\_most\_20\_members?). g2a(members). a!(self-isolate-7) \& a\textgreater a(symptoms?). (a!(self-isolate-14) \& a2g(Household). g\textgreater g(more\_than\_20\_members?). g2a(members). a!(self-isolate-7))).
This NPI consists of three NPIs that are implemented in parallel: (1) All households with up to 20 members, where a member has a positive test, go into self-isolation for 7 days, (2) Anyone with symptoms goes into self-isolation for 14 days, and (3) All households with more than 20 members, where a member experiences symptoms go into self-isolation for 7 days.
This NPI results in an average of 0.14 days of lost work during the simulated year, which is better than the baseline without an intervention (1.42 lost work days), even though the intervention from the second scenario achieved a slightly bigger decrease in the average number of lost work days (0.088 lost work days).

Finally, for scenario 4, we optimized NPIs for the combined goal function (lost\_work\_days + lost\_school\_days + 2 × sick\_days). 
The resulting intervention is: [true] a\textgreater a(positive\_test?). (a2g (Household). g2a(members). a!(self-isolate-7) \& a\textgreater a(older\_than\_67?). a\textgreater a(symp-toms?). a!(self-isolate-7)).
It tells all agents to test themselves every day and, if the test is positive, their entire household goes into self-isolation for 7 days.
The part of the NPI telling agents with positive tests to go into self-isolation for 7 days if they are older than 67 and have symptoms does not influence the NPI's effect, because the agent will be in self-isolation anyway due to the first part of the conjunction. 
The NPI results in 0.03 sick days, 0.13 lost work days, and 0.16 lost school days over the simulated year. 

\begin{figure}
    \centering
    \includegraphics[width=0.5\linewidth,trim={0cm 1.5cm 1.7cm 2cm}]{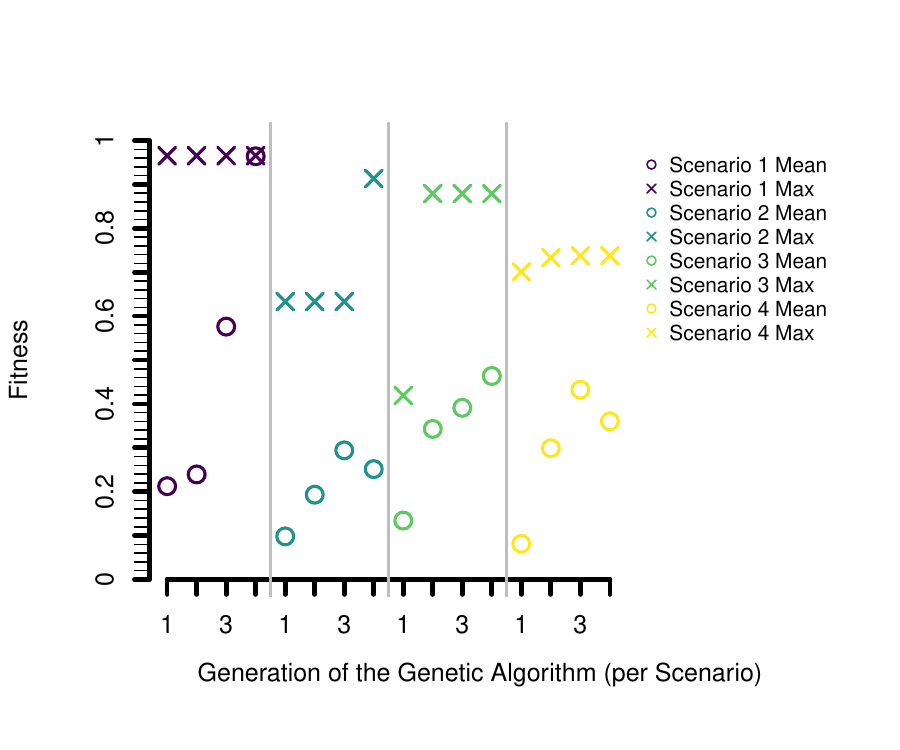}
    \caption{The mean and max fitness of the NPIs over the generations of our genetic programming algorithm per scenario (with different goal functions).
    We define the fitness to be 1/(\textless target function\textgreater  +1).
    Thus, the fitness in scenario four, where the goal function adds up the lost workdays, lost school days, and sick days, is generally lower than in the other scenarios, where only one metric makes up the goal function.
    Scenario 1: Minimize sick days; Scenario 2: Minimize lost school days per student; Scenario 3: Minimize lost work days per worker; Scenario 4: combined goal function (lost\_work\_days + lost\_school\_days + 2 * sick\_days).}
    \label{fig:caseStudy_Fitness}
\end{figure}

For all four scenarios, ADIOS found an NPI that, in a comprehensible way, reduces the targeted parameter(s) relative to the baseline simulation without NPIs.
However, the solution of the second scenario outperforms the other generated NPIs for the third and fourth scenarios.
This issue indicates that with the chosen hyperparameter configuration, ADIOS can get stuck in local maxima that depend on the population in the first generation. 

This interpretation is supported by the observation that the optimization process converged rapidly to the solutions in the third and fourth scenarios, with the final result appearing in the first or second generation for scenarios one and three (see Figure~\ref{fig:caseStudy_Fitness}).
For the first scenario, the genetic operators did not influence the finding of the resulting NPI, making ADIOS as effective as random search~\cite{koza_genetic_1994}.
However, in scenarios two and four, the final solution for the optimization problem was found in generation three (scenario four) and generation four (scenario two), indicating that the genetic operators do contribute to improving the solution candidates.

This is also supported by the observation that the population's mean fitness generally increases from one generation to the next. 
An exception can be seen in Figure~\ref{fig:caseStudy_Fitness} in scenarios two and four from generation 3 to 4, where the population's mean fitness decreases. 
This might be caused by a low locality of the mutation operator, which we discuss in more detail in Section~\ref{sec:discussion}.
But in short, by chance, there might have been mutations with large (negative) effects on the simulations' outcomes.

We also found that the runtime of experiments implementing the NPIs varies widely: we observed runtimes ranging from a few minutes to more than 10 hours for a single GEMS simulation with a generated NPI. 
Unfortunately, we did not properly record the runtimes in conjunction with the NPI being evaluated.
Therefore, we leave a more detailed analysis of correlations between runtime and NPIs to future work.

The Julia code for ADIOS and this case study can be found at \hyperlink{https://doi.org/10.5281/zenodo.19383653}{this zenodo repository}~\cite{wolpers_adios_2026}.

\section{Related Work}\label{sec:related-work}
Most existing approaches for optimizing (non-)pharmaceutical interventions in infectious disease contexts target models based on ordinary differential equations (ODEs)~\cite{lemecha_obsu_optimal_2020,perkins_optimal_2020,sharomi_optimal_2017,libin_deep_2021,colas_epidemioptim_2021,mai_planning_2023}.
The computational efficiency of ODEs relative to agent-based models makes them particularly amenable to optimization: they can serve as the foundation for Markov Decision Processes used in reinforcement learning frameworks~\cite{libin_deep_2021,mai_planning_2023,colas_epidemioptim_2021}, or be solved through mathematical methods that determine optimal enforcement levels for a predefined list of NPIs~\cite{lemecha_obsu_optimal_2020,perkins_optimal_2020,sharomi_optimal_2017}.
However, ODE models do not include the complex interactions between individuals and detailed demographic and geographical information that can be replicated using agent-based models~\cite{lorig_agent-based_2021,ponge_evaluating_2023}. 

A smaller body of work has addressed NPI optimization for agent-based simulation models by approximating intervention impacts on objective functions. 
Miikkulainen et al.~\cite{miikkulainen_prediction_2021} propose a neuroevolutionary approach that optimizes the timing and stringency of eight predefined NPIs (including school and work closures) to minimize pandemic deaths.
Their method operates on surrogate models, enabling application to both ODEs and agent-based simulations.
The Bayesian optimization method proposed by Chandak et al. also works for a range of epidemiological models, including computationally intensive agent-based models~\cite{chandak_epidemiologically_2020}.
Their approach approximates NPI effects on infections and unemployment, optimizing start times, end times, and levels of NPIs (the fraction of the population to which an NPI applies).

A limitation shared by all these methods is that they optimize only the timing (and intensity) of interventions specified a priori.
To our knowledge, no existing approach generates and evolves complex interventions as ADIOS does. 
The domain-specific language underlying ADIOS enables precise targeting of agent subpopulations based on demographic attributes, exploiting the demographic information available in agent-based models like GEMS. 
Furthermore, ADIOS offers flexibility in the specification of objective functions as any metric computable from the simulation can be incorporated into the optimization process.

\section{Discussion and Outlook}\label{sec:discussion}
While the case study above has shown that ADIOS suggests NPIs that improve selected parameters compared to a baseline simulation without NPI, we also observed that in some cases, the initially fittest solution from the first generation improves only slightly over generations.
Consequently, the quality of the final solution depends heavily on the quality of the initial population.
This begs the question whether applying the genetic algorithm is more effective than using random search over NPIs generated using our DSL and bans.
Even though our approach employs a modified variant of GGGP, there are known problems with GGGP that might relate to our observations.  

Our approach generates and optimizes NPIs represented by parse trees in our language (the genotypes), which are then mapped to experiments in a programming language (the phenotype) that are then executed to evaluate the NPI.
A popular version of GGGP is grammatical evolution (GE) that also operates on a different representation (genotype) than the representation that is executed for evaluation (phenotype)~\cite{mckay_grammar-based_2010,oneil_grammatical_2003}.
Because of differences between genotype and phenotype, the mutation in GE can have weak locality~\cite{rothlauf_locality_2006,talbi_metaheuristics_2009}.
So, small changes in the genotype can cause larger changes in the phenotype and vice versa.
In the worst case, that can result in GE performing no better than random search ~\cite{mckay_grammar-based_2010,rothlauf_locality_2006}. 

ADIOS faces a similar problem: Small mutations in the genotype's leaves can have a large impact on the NPI's effect. 
For example, in an NPI that sends all symptomatic agents into self-isolation ("[true]a\textgreater a(symptoms?).a!(self-isolate!)") only changing the leaf with the label \t{symptoms?} to \t{adult?} drastically changes the NPI's impact on the number of lost\_work\_days, as now all adults (so, most workers) are in self-isolation independently of their disease status. 
Moreover, some mutations may even have no impact on the NPIs' effect depending on the model.
For example, an NPI that sends all agents into self-isolation in conjunction with another measure will (with the language extension from the case study) always send all agents into self-isolation regardless of all other filters, mappings, and measures in the conjunction. 
Thus, the NPIs [true] (a!(self-isolate!) \& a\textgreater a(symtoms?). a!(self-isolate!)) and [true] (a!(self-isolate!) \& a\textgreater a(adult?). a!(self-isolate!)) behave identically in the simulation.
This issue might have also caused the mean fitness of the third generation in scenario four to be worse than the mean fitness of generation two in our case study (see Section~\ref{sec:caseStudy}).  

Future work could provide new insights by analyzing the impact of mutation, crossover, and other hyperparameters (i.e., population size, number of generations) for the setup used in the case study.
Additionally, examining whether certain mutations have a greater or lesser impact on the results could prove useful, and these observations could be incorporated into the mutation operator.
However, studying ADIOS in combination with the GEMS model first requires addressing the extensive runtimes of some simulations. 
We will investigate potential correlations between NPI properties and runtime, and explore ways to reduce computation time through more effective NPI formulations and parallelization based on our findings.
Furthermore, we plan to apply ADIOS to other agent-based infectious disease models to verify its flexibility and compare its performance across different models.
Finally, we aim to extend our GEMS-specific language extension to include additional trigger and measure options, enabling even more complex interventions.
This will broaden the search space and enable more exploration between generations, potentially helping to prevent local optima.

\section{Summary}\label{sec:summary}
This paper introduces ADIOS, an approach for automatically optimizing non-pharmaceutical interventions (NPIs) in agent-based infectious disease simulations. 
The challenge lies in the vast search space created by interventions that can target individuals based on multiple attributes, affect hierarchical groups, and combine arbitrarily.
ADIOS structures this search space using an adaptable domain-specific language with a context-free grammar that formalizes how triggers, filters, mappings, and measures combine.
The approach further reduces the search space through a ban mechanism that excludes semantically invalid or redundant intervention patterns. 
Grammar-guided genetic programming then evolves interventions represented as parse trees, using crossover and mutation operations while evaluating candidates through simulation.

We demonstrate ADIOS using the German Epidemic Micro-Simulation System (GEMS) across four scenarios with different optimization objectives (minimizing sick days, lost school days, lost work days, and a combined function). 
Results show that ADIOS successfully generates interventions improving targeted parameters compared to baseline simulations in three of four scenarios. 
However, rapid convergence and high sensitivity to initial populations indicate that the genetic operators may not yet provide substantial advantages over random search in this configuration, highlighting important areas for future investigation.
Future work will focus on analyzing the effect of hyperparameters, improving mutation locality, optimizing runtime performance, and validating the approach across additional agent-based models.
Although our case study identifies substantial areas for improvement, we contend that automated generation and evolution of complex interventions for agent-based epidemiological models holds promise for enhancing computational decision support.

\section*{ACKNOWLEDGMENTS}
The project „ADAPTI-M“ was funded by grants (no. 031L0322E) of the German Federal Ministry of Research, Technology and Space (BMFTR). “ADAPTI-M” is part of the Modeling Network for Severe Infectious Diseases (MONID). The author is responsible for the content of this publication.
And, this research was funded by the German Research Foundation (DFG), grant no. 320435134.

\appendix

\footnotesize
\bibliographystyle{alpha}
\bibliography{sample}

\newcommand{\etalchar}[1]{$^{#1}$}
\begin{thebibliography}{HSFTAM{\etalchar{+}}25}

\bibitem[CC20]{chatterjee_epidemics_2020}
Kaushik Chatterjee and V.~S. Chauhan.
\newblock Epidemics, quarantine and mental health.
\newblock {\em Medical Journal Armed Forces India}, 76(2):125--127, April 2020.

\bibitem[CDMK20]{chandak_epidemiologically_2020}
Amit Chandak, Debojyoti Dey, Bhaskar Mukhoty, and Purushottam Kar.
\newblock Epidemiologically and {Socio}-economically {Optimal} {Policies} via
  {Bayesian} {Optimization}.
\newblock {\em Transactions of the Indian National Academy of Engineering},
  5(2):117--127, June 2020.

\bibitem[CHR{\etalchar{+}}21]{colas_epidemioptim_2021}
Cédric Colas, Boris Hejblum, Sebastien Rouillon, Rodolphe Thiébaut,
  Pierre-Yves Oudeyer, Clément Moulin-Frier, and Mélanie Prague.
\newblock {EpidemiOptim}: {A} {Toolbox} for the {Optimization} of {Control}
  {Policies} in {Epidemiological} {Models}.
\newblock {\em Journal of Artificial Intelligence Research}, 71:479--519, July
  2021.

\bibitem[CZW{\etalchar{+}}24]{chen_metric_2024}
Wenxiu Chen, Bin Zhang, Chen Wang, Wei An, Shashika~Kumudumali Guruge, Ho-kwong
  Chui, and Min Yang.
\newblock A {Metric} of {Societal} {Burden} {Based} on {Virus} {Succession} to
  {Determine} {Economic} {Losses} and {Health} {Benefits} of {China}’s
  {Lockdown} {Policies}: {Model} {Development} and {Validation}.
\newblock {\em JMIR Public Health and Surveillance}, 10(1):e48043, June 2024.

\bibitem[dlCEdlPA05]{de_la_cruz_echeandia_attribute_2005}
Marina de~la Cruz~Echeandía, Alfonso~Ortega de~la Puente, and Manuel
  Alfonseca.
\newblock Attribute {Grammar} {Evolution}.
\newblock In José Mira and José~R. Álvarez, editors, {\em Artificial
  {Intelligence} and {Knowledge} {Engineering} {Applications}: {A}
  {Bioinspired} {Approach}}, pages 182--191, Berlin, Heidelberg, 2005.
  Springer.

\bibitem[DMNK{\etalchar{+}}18]{danandeh_mehr_genetic_2018}
Ali Danandeh~Mehr, Vahid Nourani, Ercan Kahya, Bahrudin Hrnjica, Ahmed M.~A.
  Sattar, and Zaher~Mundher Yaseen.
\newblock Genetic programming in water resources engineering: {A}
  state-of-the-art review.
\newblock {\em Journal of Hydrology}, 566:643--667, November 2018.

\bibitem[GL20]{geffen_isolation_2020}
Nathan Geffen and Marcus Low.
\newblock Isolation of infected people and their contacts is likely to be
  effective against many short-term epidemics, October 2020.

\bibitem[GSDV{\etalchar{+}}20]{giallonardo_impact_2020}
Vincenzo Giallonardo, Gaia Sampogna, Valeria Del~Vecchio, Mario Luciano,
  Umberto Albert, Claudia Carmassi, Giuseppe Carrà, Francesca Cirulli,
  Bernardo Dell’Osso, Maria~Giulia Nanni, Maurizio Pompili, Gabriele Sani,
  Alfonso Tortorella, Umberto Volpe, and Andrea Fiorillo.
\newblock The {Impact} of {Quarantine} and {Physical} {Distancing} {Following}
  {COVID}-19 on {Mental} {Health}: {Study} {Protocol} of a {Multicentric}
  {Italian} {Population} {Trial}.
\newblock {\em Frontiers in Psychiatry}, 11, June 2020.

\bibitem[HKO19]{hemberg_domain_2019}
Erik Hemberg, Jonathan Kelly, and Una-May O'Reilly.
\newblock On domain knowledge and novelty to improve program synthesis
  performance with grammatical evolution.
\newblock In {\em Proceedings of the {Genetic} and {Evolutionary} {Computation}
  {Conference}}, {GECCO} '19, pages 1039--1046, New York, NY, USA, July 2019.
  Association for Computing Machinery.

\bibitem[HSFTAM{\etalchar{+}}25]{herrera-sanchez_multiple-feature_2025}
David Herrera-Sánchez, José-Antonio Fuentes-Tomás, Héctor-Gabriel
  Acosta-Mesa, Efrén Mezura-Montes, and José-Luis Morales-Reyes.
\newblock Multiple-{Feature} {Construction} for {Image} {Segmentation} {Based}
  on {Genetic} {Programming}.
\newblock {\em Mathematical and Computational Applications}, 30(3):57, June
  2025.

\bibitem[JWL{\etalchar{+}}21]{jin_economic_2021}
Huajie Jin, Haiyin Wang, Xiao Li, Weiwei Zheng, Shanke Ye, Sheng Zhang, Jiahui
  Zhou, and Mark Pennington.
\newblock Economic burden of {COVID}-19, {China}, {January}–{March}, 2020: a
  cost-of-illness study.
\newblock {\em Bulletin of the World Health Organization}, 99(2):112--124,
  February 2021.

\bibitem[Koz93]{koza_programming_1993}
John~R. Koza.
\newblock {\em On the programming of computers by means of natural selection}.
\newblock A {Bradford} book. MIT Press, Cambridge, Mass. [u.a.], 3. print
  edition, 1993.

\bibitem[Koz94]{koza_genetic_1994}
John~R. Koza.
\newblock Genetic programming as a means for programming computers by natural
  selection.
\newblock {\em Statistics and Computing}, 4(2):87--112, June 1994.

\bibitem[KT25]{keblawi_grammatical_2025}
Mahmud Keblawi and Tomer Toledo.
\newblock Grammatical {Evolution} for automatic design of actuated traffic
  signal control plans.
\newblock {\em Applied Soft Computing}, 184:113782, December 2025.

\bibitem[LJD21]{lorig_agent-based_2021}
Fabian Lorig, Emil Johansson, and Paul Davidsson.
\newblock Agent-{Based} {Social} {Simulation} of the {Covid}-19 {Pandemic}: {A}
  {Systematic} {Review}.
\newblock {\em Journal of Artificial Societies and Social Simulation}, 24(3):5,
  2021.

\bibitem[LMV{\etalchar{+}}21]{libin_deep_2021}
Pieter J.~K. Libin, Arno Moonens, Timothy Verstraeten, Fabian Perez-Sanjines,
  Niel Hens, Philippe Lemey, and Ann Nowé.
\newblock Deep {Reinforcement} {Learning} for {Large}-{Scale} {Epidemic}
  {Control}.
\newblock In Yuxiao Dong, Georgiana Ifrim, Dunja Mladenić, Craig Saunders, and
  Sofie Van~Hoecke, editors, {\em Machine {Learning} and {Knowledge}
  {Discovery} in {Databases}. {Applied} {Data} {Science} and {Demo} {Track}},
  pages 155--170, Cham, 2021. Springer International Publishing.

\bibitem[LOFB20]{lemecha_obsu_optimal_2020}
Legesse Lemecha~Obsu and Shiferaw Feyissa~Balcha.
\newblock Optimal control strategies for the transmission risk of {COVID}-19.
\newblock {\em Journal of Biological Dynamics}, 14(1):590--607, January 2020.

\bibitem[McK01]{mckay_variants_2001}
R.~I.~(Bob) McKay.
\newblock Variants of genetic programming for species distribution modelling
  — fitness sharing, partial functions, population evaluation.
\newblock {\em Ecological Modelling}, 146(1):231--241, December 2001.

\bibitem[MFM{\etalchar{+}}21]{miikkulainen_prediction_2021}
Risto Miikkulainen, Olivier Francon, Elliot Meyerson, Xin Qiu, Darren Sargent,
  Elisa Canzani, and Babak Hodjat.
\newblock From {Prediction} to {Prescription}: {Evolutionary} {Optimization} of
  {Nonpharmaceutical} {Interventions} in the {COVID}-19 {Pandemic}.
\newblock {\em IEEE transactions on evolutionary computation}, 25(2):386--401,
  April 2021.

\bibitem[MGAS23]{mai_planning_2023}
Anh Mai, Nikunj Gupta, Azza Abouzied, and Dennis Shasha.
\newblock Planning {Multiple} {Epidemic} {Interventions} with {Reinforcement}
  {Learning}, June 2023.

\bibitem[MHW{\etalchar{+}}10]{mckay_grammar-based_2010}
Robert~I. McKay, Nguyen~Xuan Hoai, Peter~Alexander Whigham, Yin Shan, and
  Michael O’Neill.
\newblock Grammar-based {Genetic} {Programming}: a survey.
\newblock {\em Genetic Programming and Evolvable Machines}, 11(3):365--396,
  September 2010.

\bibitem[ML20]{matrajt_evaluating_2020}
Laura Matrajt and Tiffany Leung.
\newblock Evaluating the {Effectiveness} of {Social} {Distancing}
  {Interventions} to {Delay} or {Flatten} the {Epidemic} {Curve} of
  {Coronavirus} {Disease}.
\newblock {\em Emerging Infectious Diseases}, 26(8):1740--1748, August 2020.

\bibitem[M{\"o}h96]{Moehring1996}
Michael M{\"o}hring.
\newblock Social science multilevel simulation with mimose.
\newblock In Klaus~G. Troitzsch, Ulrich Mueller, G.~Nigel Gilbert, and Jim~E.
  Doran, editors, {\em Social Science Microsimulation}, pages 123--137, Berlin,
  Heidelberg, 1996. Springer Berlin Heidelberg.

\bibitem[MRU11]{maus_rule-based_2011}
Carsten Maus, Stefan Rybacki, and Adelinde~M. Uhrmacher.
\newblock Rule-based multi-level modeling of cell biological systems.
\newblock {\em BMC Systems Biology}, 5(1):166, October 2011.

\bibitem[OR03]{oneil_grammatical_2003}
Michael O’Neil and Conor Ryan.
\newblock Grammatical {Evolution}.
\newblock In Michael O’Neill and Conor Ryan, editors, {\em Grammatical
  {Evolution}: {Evolutionary} {Automatic} {Programming} in an {Arbitrary}
  {Language}}, pages 33--47. Springer US, Boston, MA, 2003.

\bibitem[PE20]{perkins_optimal_2020}
T.~Alex Perkins and Guido España.
\newblock Optimal {Control} of the {COVID}-19 {Pandemic} with
  {Non}-pharmaceutical {Interventions}.
\newblock {\em Bulletin of Mathematical Biology}, 82(9):118, October 2020.

\bibitem[PHH{\etalchar{+}}23]{ponge_evaluating_2023}
Johannes Ponge, Dennis Horstkemper, Bernd Hellingrath, Lukas Bayer, Wolfgang
  Bock, and André Karch.
\newblock Evaluating {Parallelization} {Strategies} for {Large}-{Scale}
  {Individual}-based {Infectious} {Disease} {Simulations}.
\newblock In {\em 2023 {Winter} {Simulation} {Conference} ({WSC})}, pages
  1088--1099, December 2023.

\bibitem[PLR{\etalchar{+}}20]{prem_effect_2020}
Kiesha Prem, Yang Liu, Timothy~W Russell, Adam~J Kucharski, Rosalind~M Eggo,
  Nicholas Davies, Stefan Flasche, Samuel Clifford, Carl A~B Pearson, James~D
  Munday, Sam Abbott, Hamish Gibbs, Alicia Rosello, Billy~J Quilty, Thibaut
  Jombart, Fiona Sun, Charlie Diamond, Amy Gimma, Kevin Van~Zandvoort,
  Sebastian Funk, Christopher~I Jarvis, W~John Edmunds, Nikos~I Bosse, Joel
  Hellewell, Mark Jit, and Petra Klepac.
\newblock The effect of control strategies to reduce social mixing on outcomes
  of the {COVID}-19 epidemic in {Wuhan}, {China}: a modelling study.
\newblock {\em The Lancet Public Health}, 5(5):e261--e270, May 2020.

\bibitem[PPHK25]{ponge_targeted_2025}
Johannes Ponge, Julian Patzner, Bernd Hellingrath, and André Karch.
\newblock Targeted {Household} {Quarantining}: {Enhancing} the {Efficiency} of
  {Epidemic} {Responses}.
\newblock In {\em 2025 {Winter} {Simulation} {Conference} ({WSC})}, pages
  953--964, December 2025.

\bibitem[Pro]{gesylandeu}
Geysland Project.
\newblock gesyland.eu.

\bibitem[PSHK24]{ponge_standardized_2024}
Johannes Ponge, Janik Suer, Bernd Hellingrath, and André Karch.
\newblock A {Standardized} {Framework} for {Modeling} {Non}-{Pharmaceutical}
  {Interventions} in {Individual}-{Based} {Infectious} {Disease} {Simulations}.
\newblock In {\em 2024 {Winter} {Simulation} {Conference} ({WSC})}, pages
  1106--1117, December 2024.

\bibitem[RO06]{rothlauf_locality_2006}
Franz Rothlauf and Marie Oetzel.
\newblock On the {Locality} of {Grammatical} {Evolution}.
\newblock In Pierre Collet, Marco Tomassini, Marc Ebner, Steven Gustafson, and
  Anikó Ekárt, editors, {\em Genetic {Programming}}, pages 320--330, Berlin,
  Heidelberg, 2006. Springer.

\bibitem[SM17]{sharomi_optimal_2017}
Oluwaseun Sharomi and Tufail Malik.
\newblock Optimal control in epidemiology.
\newblock {\em Annals of Operations Research}, 251(1):55--71, April 2017.

\bibitem[SSR23]{sobania_comprehensive_2023}
Dominik Sobania, Dirk Schweim, and Franz Rothlauf.
\newblock A {Comprehensive} {Survey} on {Program} {Synthesis} {With}
  {Evolutionary} {Algorithms}.
\newblock {\em IEEE Transactions on Evolutionary Computation}, 27(1):82--97,
  February 2023.

\bibitem[Tal09]{talbi_metaheuristics_2009}
El‐Ghazali Talbi.
\newblock {\em Metaheuristics: {From} {Design} to {Implementation}}.
\newblock Wiley, 1 edition, June 2009.

\bibitem[Tan07]{tanev_genetic_2007}
Ivan Tanev.
\newblock Genetic programming incorporating biased mutation for evolution and
  adaptation of {Snakebot}.
\newblock {\em Genetic Programming and Evolvable Machines}, 8(1):39--59, March
  2007.

\bibitem[W{\etalchar{+}}95]{whigham1995grammatically}
Peter~A Whigham et~al.
\newblock Grammatically-based genetic programming.
\newblock In {\em Proceedings of the workshop on genetic programming: from
  theory to real-world applications}, volume~16, pages 33--41. Tahoe City,
  California, USA, 1995.

\bibitem[WFS21]{weigl_household_2021}
Josef A.~I. Weigl, Anna-Katharina Feddersen, and Mona Stern.
\newblock Household quarantine of second degree contacts is an effective
  non-pharmaceutical intervention to prevent tertiary cases in the current
  {SARS}-{CoV} pandemic.
\newblock {\em BMC Infectious Diseases}, 21(1):1262, December 2021.

\bibitem[WPU26]{wolpers_adios_2026}
Anja Wolpers, Johannes Ponge, and Adelinde Uhrmacher.
\newblock {ADIOS} - {Case} {Study} "{Optimizing} {Interventions} for
  {Agent}-{Based} {Infectious} {Disease} {Simulations}", April 2026.

\bibitem[ZWLP22]{zhang_evaluating_2022}
Renquan Zhang, Yu~Wang, Zheng Lv, and Sen Pei.
\newblock Evaluating the impact of stay-at-home and quarantine measures on
  {COVID}-19 spread.
\newblock {\em BMC Infectious Diseases}, 22(1):648, July 2022.

\end{thebibliography}

\end{document}